\documentclass[prb, twocolumn,showpacs,preprintnumbers,amsmath,amssymb]{revtex4}

\usepackage{graphicx,color,amsmath,amssymb}

\begin{document}
\newcommand{\kk}{{\bf k}}
\newcommand{\Q}{{\bf Q}}
\newcommand{\q}{{\bf q}}
\newcommand{\gk}{g_\textbf{k}}
\newcommand{\ee}{\tilde{\epsilon}^{(1)}_\textbf{k}}
\newcommand{\HH}{\mathcal{H}}

\title{Effect of a staggered spin-orbit coupling on the occurrence of a nematic phase in Sr$_3$Ru$_2$O$_7$}

\author{Mark H. Fischer}
\author{Manfred Sigrist}
\affiliation{%
Institut f\"ur Theoretische Physik, ETH Z\"urich, 8093 Z\"urich, Switzerland
}%

\date{\today}

\begin{abstract}
Ultra-clean crystals of Sr$_3$Ru$_2$O$_7$ undergo a metamagnetic transition at low temperatures. This transition shows a strong anisotropy in the applied field direction with the critical field $H_{c}$ ranging from $\sim 5.1$T for $H$ perpendicular to $c$ to $\sim 8$T for $H\parallel c$. In addition, studies on ultra-pure samples revealed a bifurcation of the metamagnetic line for fields in $c$-direction and it has been argued that a nematic phase emerges between the magnetization jumps.
The aim of this study is to explain the field-direction anisotropy of these phenomena. Based on a microscopic tight-binding model, we introduce the metamagnetic transition by means of a van Hove singularity scenario. We show that the rotation of the O-octahedra around the $c$-axis observed in this material introduces a staggered spin-orbit coupling within the planes and naturally leads to an anisotropy in the low-temperature behavior around the metamagnetic transition. In particular, the low-temperature (nematic) phase is affected. We show that uniform in-plane magnetic fields induce a (commensurate) staggered magnetic moment component 
which can suppress the low-temperature phase. In contrast, the response to fields along the c-axis remains unaffected and thus, also the corresponding low-temperature phase . As a concrete example, we choose a nematic Pomeranchuk instability for the low-temperature phase. An experimentally testable prediction of this work is the occurrence of a staggered magnetic moment in response to
a uniform magnetic field perpendicular to the c-axis, which should be accessible by neutron scattering.
\end{abstract}

\maketitle

\section{Introduction}
\begin{figure}
	\centering
	\includegraphics{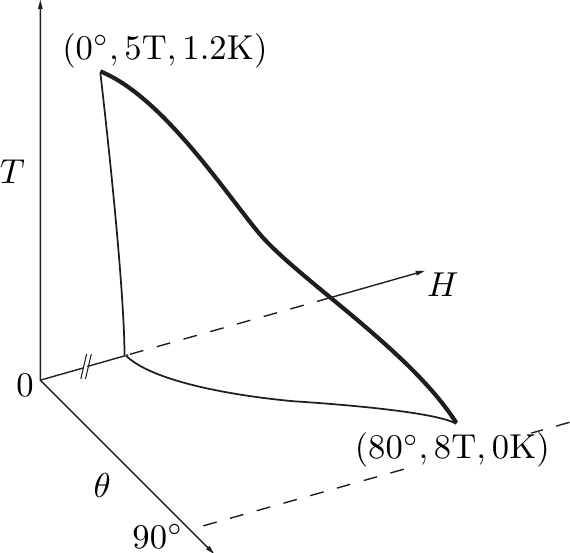}
	\caption{Schematic phase diagram of Sr$_{3}$Ru$_{2}$O$_{7}$ for fields $H$ applied with an angle $\theta$ versus the $ab$-plane. The surface represents first-order transitions separating a region with low (spin) polarization from a region with high polarization. The thick black line connecting $ (\theta,H,T) = (0^{\circ} , 5T, 1.2 K) $ and $ (80^{\circ}, 8T, 0K ) $ is a line of critical endpoints. For details see main text.}
	\label{fig:schem-phase}
\end{figure}
The ruthenium compounds of the Ruddlesden-Popper series, Sr$_{n+1}$Ru$_n$O$_{3n+1}$, have been the subject of intensive research for over a decade due to their interesting ground-states. While the quasi-two-dimensional single-layer (n=1) is an unconventional superconductor likely with $p$-wave pairing,\cite{mackenzie:2003} there is a growing tendency towards ferromagnetism with increasing layer number. The three-dimensional infinite-layer compound, SrRuO$_3$, indeed realizes an itinerant ferromagnet.~\cite{callaghan:1966} The bilayer compound (n=2) with its intermediate dimensionality, however, shows no ordering down to lowest temperatures\cite{huang:1998} and is supposed to be a strongly correlated Fermi liquid. Still, applying uniaxial pressure can induce a ferromagnetic transition and this compound is, thus, expected to be on the verge to a ferromagnetic instability.\cite{ikeda:2000} This is also supported by inelastic neutron scattering\cite{capogna:2003} and band structure calculations.~\cite{singh:2001}\\
As was first discussed by Wohlfarth and Rhodes,\cite{wohlfarth:1962} such proximity to ferromagnetism can result in metamagnetic behavior, a superlinear rise in the magnetization over a narrow region of applied magnetic field $H$. This phenomenon was observed in a number of systems~\cite{goto:2001} and also in Sr$_3$Ru$_2$O$_7$ it was found\cite{perry:2001} with a critical field $H_{c}$, however, that depends strongly on the angle of the field versus the $ab$-plane. While for $\theta =0^{\circ}$ the critical field $H_{c}\sim 5.1$T, it rises to $H_{c}\sim 8$T for $\theta =90^{\circ}$. In addition, a first-order transition occurs for in-plane fields below $T^* \approx 1.25$K, while there is only a crossover for fields parallel to $c$. It was therefore suggested that the field angle could be used as a tuning parameter for a line of first-order transitions that goes to $T^*=0$ around $\theta=80^{\circ}$, thus, realizing a quantum critical endpoint~\cite{grigera:2001} (see schematic phase diagram in Fig.~\ref{fig:schem-phase}).\\
However, when trying to reach this quantum critical endpoint on ultra-pure single crystals with residual resistivities down to $\rho_0<1\mu\Omega$cm, a splitting of the metamagnetic transition into two jumps was observed. These jumps define an intermediate phase whose exact boundaries could be determined by measuring of several thermodynamic properties.~\cite{grigera:2004} Later, it was shown that this phase breaks the symmetry of the crystal\cite{borzi:2007} and it was argued that this was due to an induced anisotropic electronic state with a symmetry-breaking Fermi surface deformation similar to a Pomeranchuck instability.~\cite{pomeranchuk:1958} That this kind of a phase, also called nematic phase in analogy to liquid crystal phases, can lead to two consecutive metamagnetic transitions had already been shown in a paper by Kee and Kim.~\cite{kee:2005}
In addition, the anomalous $T$ dependence of the susceptibility $\chi$ and the specific heat coefficient $\gamma$ could be explained.~\cite{yamase:2007b} Moreover, the two-fold degeneracy of the nematic phase allows for domain formation, such that domain-wall scattering could account for the increased resistivity of the intermediate phase.~\cite{doh:2007}\\
The electronic structure of a single layer of Sr$_3$Ru$_2$O$_7$ is dominated by bands originating from the 4d $t_{2g}$ orbitals $d_{yz}$, $d_{zx}$ and $d_{xy}$ hybridizing with the O 2p orbitals. This leads in a simple approximation to two quasi-one-dimensional bands with mainly $d_{yz}$ and $d_{zx}$ character and a two-dimensional band stemming from the $d_{xy}$ orbital. These three bands are then additionally split due to the interlayer coupling resulting in 6 bands. An important consequence of the bilayer splitting is that one of the two bands coming from the d$_{xy}$ orbitals is shifted closer to the van Hove singularity. This was also confirmed by recent angle-resolved photoemission spectroscopy (ARPES) measurements.~\cite{tamai:2008} A chemical potential in the vicinity of a van Hove singularity is the condition for the scenario described by Binz and Sigrist\cite{binz:2004} for a metamagnetic transition. Proximity to a van Hove singularity can also lead to a nematic phase accompanying a metamagnetic transition as described by Grigera et al.~\cite{grigera:2004} The anisotropy in the critical field strength could then be explained by spin-orbit coupling (SOC) effects similar to \cite{ng:2000b} leading to an anisotropic effective $g$-factor.\\ 
A different route is taken by Raghu et al.~\cite{raghu:2009} and Lee et al.~\cite{lee:09} who studied a model where the metamagnetic transition comes from the two (bilayer-split) one-dimensional bands. The anisotropy is then again introduced by considering spin-orbit interaction on the Ru-sites and the nematic phase can be understood as an orbital ordering among these one-dimensional bands.\\
Both these routes suffer, however, from a short-coming: even though they can describe the existence of a nematic phase and a dependence of $H_{c}$ on the field angle, they cannot explain why the nematic phase occurs only for fields almost parallel to the crystalline $c$-axis.\\
We will address this point in the present work. We  study a model based on a two-dimensional band in a single layer originating from the $d_{xy}$ orbitals. The bilayer effects are only taken into account by placing this band closer to the van Hove singularity. Starting from this, we will consider the effect of a lattice distortion in the planes. The O-octahedra in Sr$_3$Ru$_2$O$_7$ are rotated by $6.8^{\circ}$\cite{shaked:2000} and we will show how this introduces a staggered spin-orbit coupling, an effect similar to the Dzyaloshinski-Morya interaction for localized spins, here, however, for itinerant electrons. For magnetic fields applied in the plane, this will add a component with wave vector $\Q=(\pi, \pi)$ to the static susceptibility. The induced commensurate spin-density wave (SDW) will open gaps in the Fermi surface close to the van Hove points which will have an impact on the occurrence of any instability that emerges due to the proximity to a van Hove singularity. We choose here the electronic nematic phase to examine this aspect, since it relies on the presence of the van Hove singularity and represents one of  the most promising candidates for the intermediate phase. In order to discuss the essential influence of the spin-orbit coupling on the phase diagram we adopt here a mean-field approach, with the short-coming that critical fluctuations are not included well. While, in particular, quantum critical fluctuations represent an intriguing part of the phenomenology of this metamagnetic transition, we assume that they are not essential to understand the basic effects due to spin-orbit coupling and lie beyond the scope of this study.\\
This paper is organized as follows: in section II, we will introduce our model based on a three-band Hamiltonian consisting of the Ru 4$d_{xy}$ orbital and the in-plane O 2$p$ orbitals. After reducing this model to an effective one-band model, we will analyze the effect of a rotation of the oxygen octahedra on it. On-site interactions are then treated within mean-field theory and an additional applied magnetic field is considered. The resulting model is then studied in a next section. Following Metzner et al.,\cite{metzner:2003} we will in section IV add a forward-scattering term to allow for a nematic phase and study the influence of the staggered spin-orbit coupling to this phase. In a last section we will discuss and summarize our findings. 

\begin{figure}
	\centering
	\includegraphics{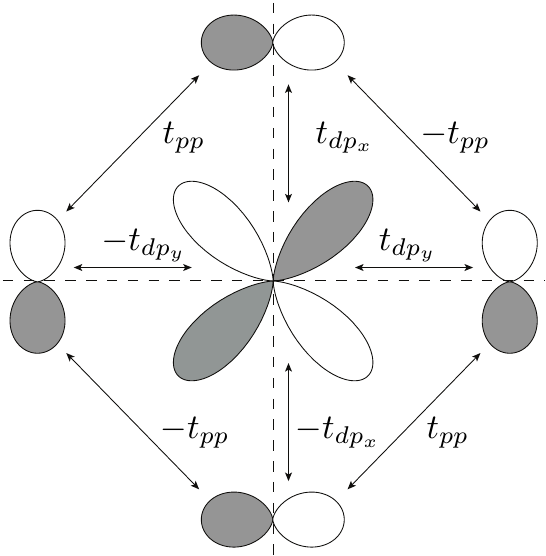}
	\caption{Different hoppings in the three-band model of Ru 4 d$_{xy}$ and O 2 p$_x$ / p$_y$ orbitals. The relative signs of the hopping integrals come from the phases of the orbital wave-functions.}
	\label{fig:hopping1}
\end{figure}
\section{Model}
\subsection{Basic Hopping Hamiltonian}
The starting point for our model is a three-band tight-binding model including the in-plane 4d Ru orbital (d$_{xy}$) and two 2p O orbitals (p$_x$ and p$_y$) with on-site energies $E_d$ and $E_p=E_d - \Delta$, respectively. In this model, an electron can hop from a d$_{xy}$ orbital in $x$ ($y$) direction to a p$_y$ (p$_x$) oxygen orbital and vice versa with the hopping integral $t_{dp_y}$ ($t_{dp_x}$). Additionally, due to strong hybridization of the oxygen 2p orbitals, electrons can hop between neighboring oxygen orbitals. This leads to a Hamiltonian of the form
\begin{equation}
	\HH_{3b} = \sum_{s}\vec{C}^{\dag}_{s}\left(\begin{array}{ccc}E_{d}&\tilde t_{dp_{x}}&\tilde t_{dp_{y}}\\\tilde t_{dp_{x}}&E_{p_{x}}&\tilde t_{pp}\\\tilde t_{dp_{y}}&\tilde t_{pp}&E_{p_{x}}\end{array}\right)\vec{C}^{\phantom{\dag}}_{ s}
\end{equation}
where $\vec{C}^{\dag}=(d^{\dag}, p^{\dag}_{x}, p^{\dag}_{y})$ are the creation operators for the above mentioned orbitals. Care has to be taken of the different signs of the hopping integrals due to the phase of the orbital wave-functions indicated by the tildes (see Fig.~{\ref{fig:hopping1}}).\\
To integrate out the high-energy degrees of freedom and thus, to reduce our model to one band, we construct an effective Hamiltonian only living at the Ru sites,\cite{noce:97, cuoco:02}
\begin{equation}\label{eq:const-eff}
	\HH_{eff} = \sum_{p}\frac{\HH_{3b}|p\rangle\langle p|\HH_{3b}}{E_d - E_p} + \sum_{pp'}\frac{\HH_{3b}|p\rangle\langle p|\HH_{3b}|p'\rangle\langle p'|\HH_{3b}}{(E_d - E_p)(E_d - E_{p'})}
\end{equation}
where the sums run over all oxygen orbitals on all sites. This leads to a simple hopping Hamiltonian,
\begin{equation}
	\HH^{(0)} = - t \sum_{\langle i, j\rangle}\sum_s c^{\dag}_{is}c^{\phantom{\dag}}_{js} - t' \sum_{(i,j)}\sum_s c^{\dag}_{is}c^{\phantom{\dag}}_{js},
\end{equation}
where $c^{\dag}_{is}$ creates an electron at Ru site $i$ with spin $s$, $\langle i,j\rangle$ denotes nearest neighbors (nn) and $(i,j)$ next-nearest neighbors (nnn). The hopping integrals in this effective Hamiltonian then read to lowest order
\begin{equation}
	t = \frac{t_{dp}^2}{\Delta}\quad {\rm and} \quad t' = \frac{t_{dp}^2t^{\phantom 2}_{pp}}{\Delta^2}
\end{equation}
with $t_{dp_x}=t_{dp_y}=t_{dp}$.

\begin{figure}
	\centering
	\includegraphics{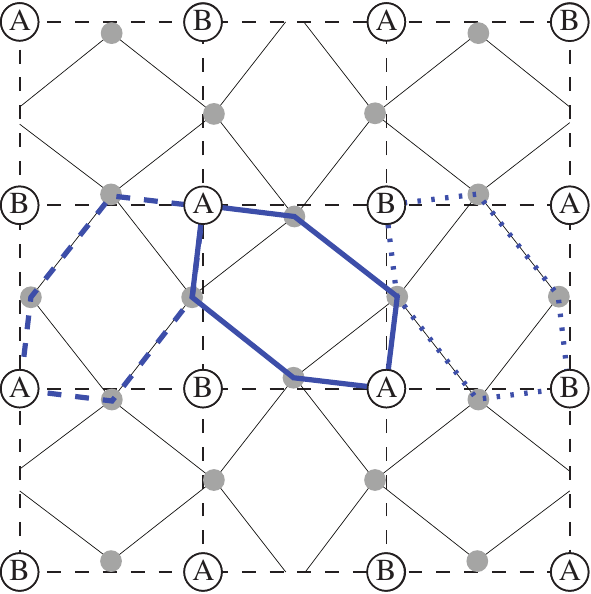}
	\caption{The RuO$_2$ plane with rotated oxygen octahedra leading to a doubling of the unit cell. The two distinct lattice sites are denoted as $A$ and $B$. The inversion symmetry of the bonds between Ru-ions is broken leading to a staggered spin-dependent nearest-neighbor hopping. The next-nearest-neighbor hopping is still isotropic and the same for both lattice sites as is depicted by the different, equivalent hopping-paths (bold solid, dashed lines for $A$ to $A$ and dotted line for $B$ to $B$).}
	\label{fig:rotated}
\end{figure}
\begin{figure}
	\centering
	\includegraphics{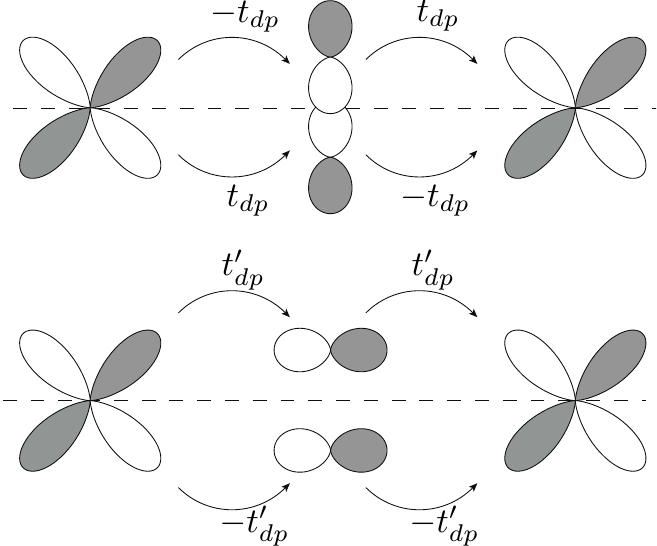}
	\caption{Possible hoppings for the three-band model in the case of rotated O$_{6}$ octahedra. The sign change in the hopping integrals is due to the phase of the Wannier functions and holds in first order.}
	\label{fig:hopping2}
\end{figure}

If we additionally take the rotated oxygen-octahedra into account, we obtain a bipartite lattice leading to a $\sqrt{2}\times\sqrt{2}$ larger unit cell separating A and B sublattices (see Fig.~\ref{fig:rotated}). A consequence is that the formerly symmetry-forbidden hopping in $x$- ($y$-) direction via p$_x$ (p$_y$) orbitals is now possible with matrix element $t_{dp}'$ as is depicted in Fig.~\ref{fig:hopping2}. Thus, we include spin-orbit coupling at the oxygen site, $\HH^{O-2p} = \lambda L_z S_z$, which mixes the $p_x$ and $p_y$ orbitals. We therefore need to change to eigenfunctions of the spin-orbit coupling, $|\pm\rangle$, with $\HH^{O-2p}|\pm\rangle = \pm \lambda s|\pm\rangle$, $s=\pm1$ the spin index.\\
As an example, we write the total Hamiltonian for the sublattice A for the $x$-direction,
\begin{eqnarray}
	\HH^{(x)}_A &=& - \sum_{j \in A}\sum_s(\tilde{t} d^{\dag}_{js}p^{\phantom{\dag}}_{+, j+\hat{x}/2s} + \tilde{t}^*d^{\dag}_{js}p^{\phantom{\dag}}_{+, j-\hat{x}/2s}+ {\rm h.c.})\nonumber\\
	&& - \sum_{j \in A}\sum_s(\tilde{t}^*d^{\dag}_{js}p^{\phantom{\dag}}_{-, j+\hat{x}/2s} + \tilde{t}d^{\dag}_{js}p^{\phantom{\dag}}_{-, j-\hat{x}/2s}+ {\rm h.c.})\nonumber\\
	&& - \sum_{\nu=\pm}\sum_{as}(\Delta \pm \lambda s)p^{\dag}_{\nu,as}p^{\phantom{\dag}}_{\nu, as},
\end{eqnarray}
where $p^{\dag}_{\pm, j+\hat{x}/2s}$ creates an electron at the oxygen site $j+\hat{x}/2$ in the $|\pm\rangle$ state  with spin $s$ and 
\begin{equation}
	\tilde{t} = \frac{t'_{dp} - i t_{dp}}{\sqrt{2}}.
\end{equation}
Applying perturbation theory in the from of Eq.~\eqref{eq:const-eff} to this Hamiltonian to construct an effective model, we find for the hopping integral from a site $A$ to a site $B$ in the positive $x$-direction
\begin{eqnarray}\label{eq:rot_hop}
	\langle A |\HH_{eff}|B\rangle &=& \frac{\tilde{t}^{2}}{\Delta + \lambda s} + \frac{(\tilde{t}^{*})^{2}}{\Delta - \lambda s}\nonumber\\
	&=& -(t_{pd}^2 - t_{pd}'^2)\frac{\Delta}{\Delta^2 - \lambda^2} + i s \frac{2\lambda t_{pd}t_{pd}'}{\Delta^2-\lambda^2}\nonumber\\
	&=& -t + i \alpha s.\label{eq:hops}
\end{eqnarray}
Hence, we have a total Hamiltonian $\HH = \HH^{(0)} + \HH^{soc}$ with a nn-and nnn-hopping Hamiltonian $\HH^{(0)}$ and a staggered spin-dependent hopping with the form of a staggered SOC of Rashba-type,
\begin{eqnarray}
	\HH^{soc} \!=\! \sum_{s s'}\!\Big[\!\!\!&-&\!\!\!i\alpha\! \sum_{j\in A}\sum_{\hat{a} = \hat{x}, \hat{y}}\!(c^{\dag}_{j + \hat{a} s}c^{\phantom{\dag}}_{j s'} - \!c^{\dag}_{j s'}c^{\phantom{\dag}}_{j + \hat{a} s})\sigma^{z}_{ss'}\nonumber\\
	\!\!\!&+&\!\!\!i\alpha\! \sum_{j\in B}\sum_{\hat{a} = \hat{x}, \hat{y}}\!(c^{\dag}_{j + \hat{a} s}c^{\phantom{\dag}}_{j s'} - \!c^{\dag}_{j s'}c^{\phantom{\dag}}_{j + \hat{a} s})\sigma^{z}_{ss'}\!\Big].
\end{eqnarray}
Note that the nnn-hopping integrals, even though renormalized, do not become anisotropic, which can be deduced from geometrical considerations as indicated in Fig.~\ref{fig:rotated}.\\
The bipartite lattice introduces a wave-vector $\Q=(\pi,\pi)$ with which the total Hamiltonian in momentum space reads
\begin{equation}
	\HH = \frac{1}{2}\sum_{s s'}\sum_{\kk}\!\phantom{}^{'}\vec{c}^{\dag}_{\kk  s}\!\!\left(\!\!\begin{array}{cc} \varepsilon^{(1)}_{\kk} + \varepsilon^{(2)}_{\kk} - \mu & ig_{\kk}\sigma_{s s'}^{z} \\ -ig_{\kk}\sigma_{s s'}^{z} & -\varepsilon^{(1)}_{\kk} +\varepsilon^{(2)}_{\kk} - \mu\end{array}\!\!\right)\!\! \vec{c}^{\phantom{\dag}}_{\kk  s'}
\end{equation}
where $\vec{c}^{\dag}_{\kk s}=(c^{\dag}_{\kk  s},c^{\dag}_{\kk+\Q  s})$. Here, $\varepsilon_{\kk} = -2t(\cos k_x + \cos k_y) - 4t'\cos k_x \cos k_y = \varepsilon^{(1)}_{\kk} + \varepsilon^{(2)}_{\kk}$ are the hopping energies for nn- and nnn-hopping and $g_{\kk} = 2\alpha(\cos k_x + \cos k_y)$ is the form factor for the spin-orbit coupling. The prime in the $\bf{k}$- summation indicates that the summation runs only over the reduced first Brillouin zone. As this restriction will hold for all subsequent $\bf{k}$-sums we will omit the prime in the following.\\
The staggered SOC hybridizes states with $\kk$ and $\kk + \Q$. It is now convenient to introduce Pauli matrices $(\tau^0, \vec{\tau})$ in momentum space $\{\kk, \kk + \Q\}$ such that we can write the Hamiltonian as
\begin{equation}\label{eq:h0}
	\mathcal{H}=\sum_{\kk s s'}\vec{c}^{\dag}_{\kk s}\mathcal{H}^{\phantom{\dag}}_{\kk ss'}\vec{c}^{\phantom{\dag}}_{\kk s'}
\end{equation}
with
\begin{equation}\label{eq:ham0}
	\mathcal{H}^{(0)}_{\kk ss'} = (\varepsilon_{\kk}^{(2)} - \mu)\sigma_{ss'}^{0}\tau^0+\varepsilon_{\kk}^{(1)}\sigma^0_{ss'}\tau^z - g_{\kk}\sigma^z_{ss'}\tau^y.
\end{equation}
Diagonalizing this Hamiltonian yields two (spin-degenerate) bands,
\begin{equation}
	\xi_{\alpha\kk s} = \epsilon^{(2)}_{\kk} + (-1)^{\alpha} \sqrt{(\epsilon^{(1)}_{\kk})^2 + g_{\kk}^2}
\end{equation}
where $\alpha=1,2$. The first Brillouin zone is folded back, as can be seen in Fig.~\ref{fig:fs}a, where the Fermi surface is plotted. This is in accordance with the doubling of the unit cell introduced by the rotated oxygen octahedra.
\subsection{Magnetization and On-site Interaction}
Before adding an on-site interaction to the Hamiltonian \eqref{eq:h0}, we first want to examine the effect of an applied magnetic field. From the form of Eq.~\eqref{eq:ham0} we first see that a magnetic field in $z$-direction is a mere spin-dependent shift of the chemical potential. The response of the system is thus a simple polarization. However, for in-plane fields the staggered spin-orbit coupling introduces a coupling of homogenous magnetic fields to a staggered magnetization, i.e. a commensurate SDW. This means that the static spin-susceptibility has a component with wave-vector $\Q$.\\
To see this, we add a Zeeman term of the form
\begin{equation}
	\HH^Z = g \mu_B[ \vec{H}_0 \cdot\vec{S}(0) +  \vec{H}_{\bf Q}\cdot \vec{S}({\bf Q})]
\end{equation}
with the spin operators 
\begin{equation}
	\vec{S}({\bf q}) = \frac{1}{2}\sum_{\kk} c^{\dag}_{\kk + {\bf q} s}\vec{\sigma}_{ss'}c^{\phantom{\dag}}_{\kk s'}=\sum_{\kk}\vec{S}_{\kk}({\bf q}),
\end{equation}
$\mu_B$ the Bohr magneton and the Land\'e factor $g$. This corresponds to a homogenous and a staggered magnetic field, in accordance with the structure of the Hamiltonian given in eq.~(\ref{eq:ham0}) and can, thus, be written  as 
\begin{equation}\label{eq:zeeman}
	\mathcal{H}^{Z}_{\kk ss'} = (\vec{h}_0\cdot\vec{\sigma}\tau^0 + \vec{h}_{\bf Q}\cdot\vec{\sigma}\tau^x).
\end{equation}
Here, we have introduced $\vec{h}_{0/\vec{Q}}=\vec{H}_{0/\vec{Q}} / H_{0}$ where $H_{0}=2\cdot 10^{-4}t / (g \mu_{B})$.
It is now straightforward to calculate the magnetic response of the system to an applied field by using the thermodynamic relation
\begin{equation}
	\langle m^{i}_{0/\bf Q}\rangle = -\frac{\partial}{\partial h^{i}_{0/\bf Q}} F(T, \vec{h}_0, \vec{h}_{\bf Q}, N)
\end{equation}
where $\langle m^{i}_0\rangle$ is the homogenous magnetization pointing in the $i$-direction while $\langle m^{i}_{\bf Q}\rangle$ corresponds to a staggered magnetization. Using \eqref{eq:zeeman} together with \eqref{eq:h0}, we find for the case of a homogeneous field in $x$-direction a finite staggered magnetization in $y$-direction,
\begin{equation}
	\langle m^{y}_{\Q}\rangle = \sum_{\alpha=1,2}\sum_{\beta=\pm}\sum_{\kk}n_F(\xi_{\alpha\beta,\kk})\frac{(-1)^\alpha g_{\kk}}{\sqrt{(h^x_0\pm\epsilon^{(1)}_{\kk})^2 + g_{\kk}^2}}.
\end{equation}
In the above equation, $\xi_{\alpha\pm,\kk} = \epsilon^{(2)}_{\kk} + (-1)^{\alpha} \sqrt{(h^x_0\pm\epsilon^{(1)}_{\kk})^2 + g_{\kk}^2}$ are the four energy bands in a homogenous field in $x$-direction and $n_F(\xi)$ is the Fermi distribution function.\\

If we introduce an on-site interaction term to the Hamiltonian,
\begin{equation}\label{eq:U}
	\HH^{U} = U\sum_{i}n^{\vec{a}}_{i\uparrow}n^{\vec{a}}_{i\downarrow}
\end{equation}
which we want to treat within mean-field theory, we first have to choose an appropriate spin-quantization axis (indicated by the superscript $\vec{a}$). For simplicity, we only consider the two cases of an applied magnetic field in $z$- and in $x$-direction.
\subsubsection{Field applied in $z$-direction}
Since a field perpendicular to the plane does not couple to any staggered magnetization, the first case is straightforward. The quantization axis is the $z$-axis and we write 
\begin{equation}
	\HH^{U} = U\sum_{i}\Big[\frac{(n_{i\uparrow}+n_{i\downarrow})^2}{4} - \frac{(n_{i\uparrow}-n_{i\downarrow})^2}{4}\Big] .
\end{equation}
Since we do not expect large fluctuations in the charge density, the first term is a constant and, thus, the interaction can be written as
\begin{equation}\label{eq:spinspin}
	\HH^{U} = -U\sum_{i}S^z_iS^z_i
\end{equation}
with $S^z_i = (n_{i\uparrow}-n_{i\downarrow})/2$. Applying mean-field theory to this expression and changing to momentum space yields
\begin{equation}
	\HH^{U} = -2 U M^{z} \sum_{\kk}S^z_{\kk}(0) + U N (M^z)^2
\end{equation}
with $M^{z}=\langle S^z_i\rangle$ independent of site $i$. Therefore, this leads to an additional term in (\ref{eq:ham0})
\begin{equation}
	\HH^{U}_{\kk} = - U M^{z} \sigma^z\tau^0 + U  (M^z)^2.
\end{equation}
The total Hamiltonian now reads
\begin{equation}\label{eq:hamz}
	\mathcal{H}_{\kk ss'} = \mathcal{H}^{(0)}_{\kk ss'} + \tilde{h}_0^z\sigma^z\tau^0
\end{equation}
with the effective magnetic field $\tilde{h}_0^z = h_0^z - U M^{z}$ for simplicity. This Hamiltonian has four eigenenergies
\begin{equation}\label{eq:xi-z}
    \xi_{\alpha\kk s} = \epsilon^{(2)}_\textbf{k} + s \tilde{h}_0^z + (-1)^{\alpha} \sqrt{g_\textbf{k}^2 + (\epsilon^{(1)}_\textbf{k})^2}-\mu
\end{equation}
and via the grand-canonical potential per lattice site,
\begin{equation}
	\omega = -T\sum_{\alpha = 1,2}\sum_{\kk s}\log[1 + \exp(-\xi_{\alpha\kk s}/T)] + U (M^{z})^2,
\end{equation}
the self-consistency equations can be derived,
\begin{eqnarray}
	n &=& \frac{1}{N}\sum_{\alpha =1,2}\sum_{\kk s}n_F(\xi_{\alpha\kk s})\label{eq:zn},\\
	M^{z} &=& \frac{1}{2N}\sum_{\alpha = 1,2}\sum_{\kk s} s\, n_F(\xi_{\alpha\kk s}).\label{eq:zmag}
\end{eqnarray}
\begin{figure}
	\centering
	\includegraphics{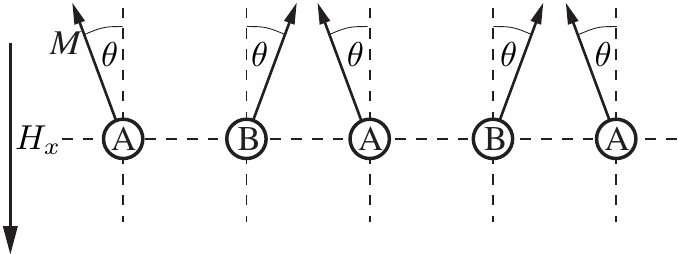}
	\caption{Real space schematic of the magnetic order for the case of an applied field in the $xy$-plane: Due to the staggered spin-orbit coupling, the magnetization is also staggered with respect to the sublattice sites A and B with order parameter $M$ and $\theta$ for the total moment and canting angle, respectively.}
	\label{fig:canting}
\end{figure}
\begin{figure}
	\centering
	\includegraphics{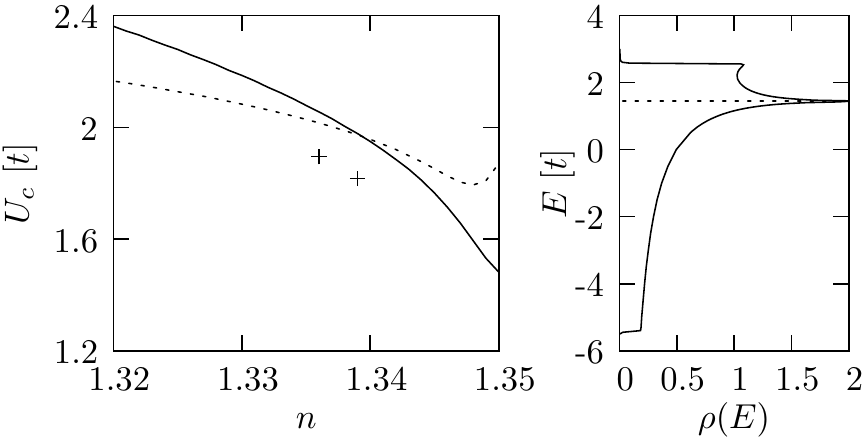}
	\caption{Left Figure: Critical interaction strength for the ferromagnetic (solid line) and the SDW instability (dashed line). The crosses denote the choices for the interaction strength used for the two densities in Fig.~\ref{fig:Hc}. Right Figure: Density of states in the absence of an external field showing the van Hove singularity at $\mu=4t'=1.44t$ (dashed line).}
	\label{fig:Uc}
\end{figure}
\subsubsection{Field applied in $x$-direction}
The case of  a field in $x$-direction is slightly more involved since we expect an additional response in form of a staggered magnetization in $y$-direction leading to a canted magnetization in $x$-direction. Therefore, we define an angle $\theta$ for the canting angle and denote with $M$ the total moment (see Figure~\ref{fig:canting}). The quantization axis, $\hat{a}$ and $\hat{b}$ for the spin on lattice sites $A$ and $B$, respectively, in Eq.~\eqref{eq:U} should therefore be perpendicular to $z$ and tilted away from the $x$-direction by the angle $\theta$. We start from (\ref{eq:spinspin}) and decouple to find
\begin{eqnarray}
	\HH^U &=& -U\sum_{i\in A}S_i^{\hat{a}}S_i^{\hat{a}} - U\sum_{i\in B}S_i^{\hat{b}}S_i^{\hat{b}}\nonumber\\
	&=& U N M^2 - 2 U M \sum_{i\in A}(S^x_i\cos\theta +S^y_i\sin\theta )\nonumber\\
	&& - 2 U M \sum_{i\in B}(S^x_i\cos\theta -S^y_i\sin\theta). 
\end{eqnarray}
Changing again to momentum space, this results in an additional term in the total Hamiltonian
\begin{equation}
	\HH^{U}_{\kk} = -(U M \cos\theta) \sigma^x\tau^0 - (UM \sin\theta)\sigma^y\tau^x+ U N M^2
\end{equation}
which then becomes
\begin{equation}
	\mathcal{H}_{\kk ss'} = \mathcal{H}^{(0)}_{\kk ss'} + \tilde{h}_0^x\sigma^x\tau^0 +\tilde{m}^y\sigma^y\tau^x
\end{equation}
with $\tilde{m}^y = -UM\sin\theta$ and
\begin{equation}
	\tilde{h}_0^x = h_0^z -UM\cos\theta.
\end{equation}
The eigenvalues read
\begin{eqnarray*}
    \xi_{1\pm,\textbf{k}} &=& \epsilon^{(2)}_\textbf{k} - \sqrt{(\tilde{m}^y\pm g_{\kk})^2 + (\tilde{h}^x_0\pm\epsilon^{(1)}_{\kk})^2}-\mu,\\
    \xi_{2\pm,\textbf{k}} &=& \epsilon^{(2)}_\textbf{k} + \sqrt{(\tilde{m}^y\pm g_{\kk})^2 + (\tilde{h}^x_0\pm\epsilon^{(1)}_{\kk})^2}-\mu.
\end{eqnarray*}
Again, the self-consistency equations can be deduced from the grand-canonical potential and are
\begin{eqnarray}
	n &=& \frac{1}{N}\sum_{\alpha}\sum_{\beta=\pm}\sum_{\kk}n_F(\xi_{\alpha\beta,\kk}),\\
	0 &=& \frac{UM}{N}\sum_{\alpha}\sum_{\beta=\pm}\sum_{\kk}n_F(\xi_{\alpha\beta,\kk})\frac{\partial \xi_{\alpha\beta,\kk}}{\partial \theta},\label{eq:sc-theta}\\
	M &=& \frac{1}{2N}\sum_{\alpha}\sum_{\beta=\pm}\sum_{\kk}n_F(\xi_{\alpha\beta,\kk}) \frac{\partial \xi_{\alpha\beta,\kk}}{\partial M}\label{eq:xymag}
\end{eqnarray}
with
\begin{eqnarray}
\!\!\frac{\partial \xi_{\alpha\beta,\kk}}{\partial \theta}\!\!&=&\!\! (\!-1\!)^\alpha\frac{\mp g_{\kk}\cos\theta - (h_0^x/2 \mp  \epsilon^{(1)}_{\kk})\sin\theta}{\sqrt{(\tilde{m}^y\pm g_{\kk})^2 + (\tilde{h}^x_0\pm\epsilon^{(1)}_{\kk})^2}}\label{eq:xi-theta},\\
\!\!\frac{\partial \xi_{\alpha\beta,\kk}}{\partial M}\!\!&= &\!\! (\!-1\!)^\alpha\frac{(h_0^x/2\!\pm\! \epsilon^{(1)}_{\kk})\cos\theta\! -\! UM\! \pm\! g_{\kk}\sin\theta}{\sqrt{(\tilde{m}^y\pm g_{\kk})^2 + (\tilde{h}^x_0\pm\epsilon^{(1)}_{\kk})^2}}.
\end{eqnarray}

From Eqs.~(\ref{eq:sc-theta}) and (\ref{eq:xi-theta}) it follows trivially that there is only a canted magnetization if there is a finite SOC.\\
\section{results}
\subsection{Field in $z$-direction}
\begin{figure}
	\centering
	\includegraphics{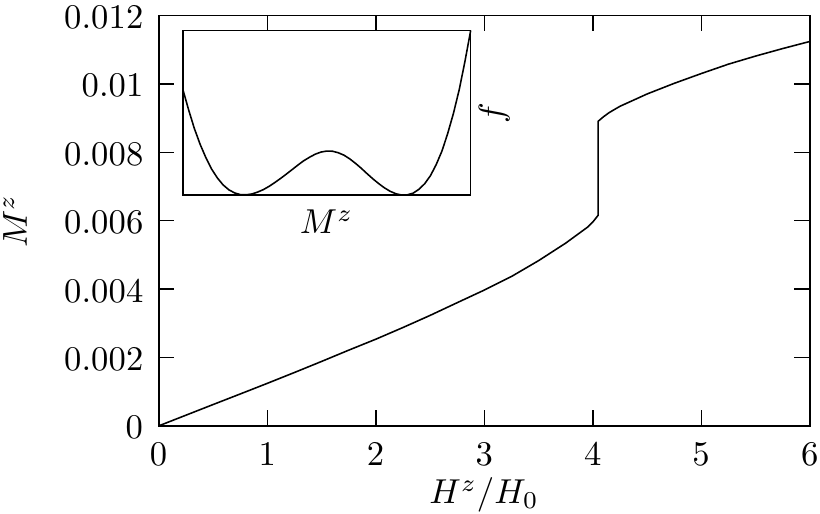}
	\caption{Magnetization for $n=1.336$ for an applied field in $z$-direction for a temperature of $T=5\cdot10^{-4}t$. Inset: the free energy $f$ for $H=H_c$ as a function of magnetization between $M^z=0.005$ and $0.01$ to emphasize the first-order character of the metamagnetic transition.}
	\label{fig:z-meta}
\end{figure}
Since a metamagnetic transition is expected to occur close to magnetic instabilities, we investigate their occurrence from Eq.~(\ref{eq:zmag}). The linearized self-consistent equation yields the condition
\begin{equation}\label{eq:z-cond}
	1 - \frac{U}{N}\sum_{\alpha\kk }\frac{1}{4T \cosh^2(\xi_{\alpha\kk s}/2T)} =0 
\end{equation}
for the occurrence of a ferromagnetic instability, the familiar Stoner criterion. This is not surprising since, apart from the folding of the Brillouin zone, the staggered spin-orbit coupling in the case of a magnetic field in $z$-direction only leads to a renormalization of the nearest-neighbor hopping (c.f. Eq.~(\ref{eq:xi-z})). The critical interaction strength for a ferromagnetic instability to occur as a function of the electron density, $n$, is shown in the left part of Fig.~\ref{fig:Uc} (solid line). It drops significantly close to a density of $n_{vH}\approx1.35$. This corresponds to $\mu_{vH}=4t'$ where the Fermi surface hits the van Hove points located at $(\pm\pi, 0)$ and $(0, \pm\pi)$ thus leading to a diverging density of states. This divergence is shown in the right part of Fig.~\ref{fig:Uc}, where the density of states in zero field is plotted.\\
For a further analysis of the metamagnetic transition with applied field in $z$-direction, we fix the density of electrons slightly below $n_{vH}$, $n=1.336$, and choose an interaction strength $U$ close to the critical one obtained from the linearized self-consistency equation (see Fig.~\ref{fig:Uc}). Here and in following numerical calculations, we keep the spin-orbit coupling strength at $\alpha = 0.05t$. The magnetization curve obtained is shown in Fig.~\ref{fig:z-meta}. To emphasize the first-order nature of the transition at $T=5\cdot10^{-4}t$, the inset shows the free energy at the critical magnetic field as a function of magnetization $M^z$ for values between $0.005$ and $0.01$.\\
\begin{figure}
	\centering
	\includegraphics{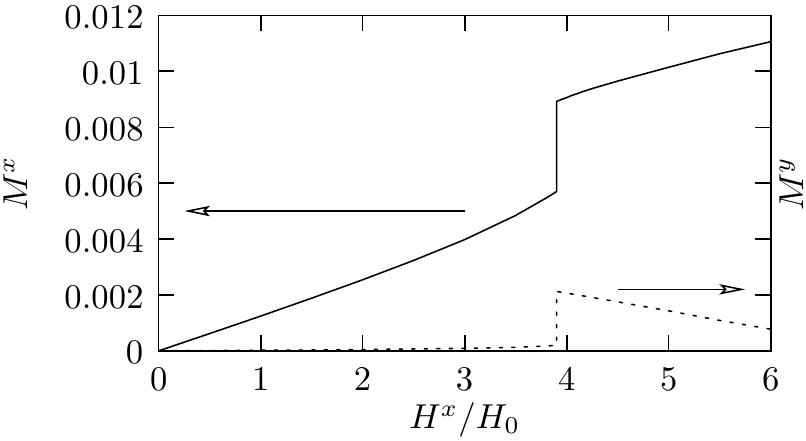}
	\caption{Uniform magnetization $M^x$ (solid line) and staggered magnetization $M^y$ (dashed line) for $n=1.336$ for an applied field in $x$-direction for a temperature of $T=5\cdot10^{-4}t$ and $\alpha=0.05t$. For this temperature, the metamagnetic transition is clearly of first order.}
	\label{fig:xy-meta}
\end{figure}
\subsection{Field in $x$-direction}
For the case of the field applied in $x$-direction, we can again explore the occurrence of a magnetic instability by linearizing the self-consistency equation~\eqref{eq:xymag}, leading to
\begin{widetext}
\begin{equation}\label{eq:m-lin}
	1-\frac{U}{N}\Big[\sum_{\alpha\beta}\sum_{\kk}\frac{1}{4T \cosh^2(\xi_{\alpha\beta,\kk}/2T)}\frac{(g_{\kk}\sin\theta+ \epsilon^{(1)}_{\kk}\cos\theta)^2}{g_{\kk}^2 + (\epsilon^{(1)}_{\kk})^2}+\sum_{\alpha\beta}\sum_{\kk}n_F(\xi_{\alpha\beta,\kk}) (-1)^\alpha\frac{(g_{\kk}\cos\theta- \epsilon^{(1)}_{\kk}\sin\theta)^2}{[g_{\kk}^2 + (\epsilon^{(1)}_{\kk})^2]^{3/2}} \Big] =0.
\end{equation}
\end{widetext}
To further analyze this, it is useful to change to different order parameters, from ($M, \theta$) to ($M^x=M\cos\theta, M^y=M\sin\theta$) with $M^{x}$ the uniform magnetization in $x$-direction and $M^{y}$ the staggered component in $y$-direction, respectively. Writing the self-consistency equations in these new parameters,
\begin{eqnarray}
	M^x\!\!&=&\!\!\frac{1}{2N}\!\!\sum_{\alpha,\beta}\!\!\sum_{\kk}n_F(\xi_{\alpha\beta,\kk}) \frac{(-1)^\alpha(\pm \epsilon^{(1)}_{\kk} - UM^x)}{\sqrt{\!(\tilde{m}^y\pm g_{\kk})^2\! +\! (\tilde{h}^x_0\pm\epsilon^{(1)}_{\kk})^2\!}},\nonumber\\
	M^y\!\!&=&\!\!\frac{1}{2N}\!\!\sum_{\alpha,\beta}\!\!\sum_{\kk}n_F(\xi_{\alpha\beta,\kk})\frac{ (-1)^\alpha(\pm g_{\kk} - UM^y)}{\sqrt{\!(\tilde{m}^y\pm g_{\kk})^2\! +\! (\tilde{h}^x_0\pm\epsilon^{(1)}_{\kk})^2\!}}\nonumber
\end{eqnarray}
and linearizing this system of equations, 
\begin{equation}
	\left(\begin{array}{c}M^x\\M^y \end{array}\right)=\left.\left(\begin{array}{cc}\partial_{x}M^x & \partial_{y}M^x\\ \partial_{x}M^y & \partial_{y}M^y\end{array}\right)\right|_{M^x=M^y=0}\left(\begin{array}{c}M^x\\M^y\end{array}\right)
\end{equation}
leads to two different possible magnetic instabilities, 
\begin{eqnarray}
	0 &=& 1 -  \frac{U}{N}\sum_{\alpha\textbf{k}} \frac{1}{4T\cosh^2(\xi_{\alpha\beta, \kk}/2T)},\\
	0 &=& \sqrt{t^2 + \alpha^2} +  \frac{U}{N}\sum_{\alpha\textbf{k}}\frac{(-1)^\alpha n_F(\xi_{\alpha\beta,\kk})}{2|\cos k_x + \cos k_y|}.
\end{eqnarray}
The first one is the same as Eq.~\eqref{eq:z-cond} and corresponds to a ferromagnetic instability. The second equation corresponds to a SDW instability occurring due to the near nesting of the Fermi surfaces. Note, however, that in both cases, the magnetization will have a uniform as well as a staggered component.\\
The solutions of these equations as functions of the electron filling, $n$, are plotted in Fig.~\ref{fig:Uc}. We see that the critical interaction strength for the spin-density wave instability is generally below the one for a ferromagnetic instability, but shoots up when approaching the critical filling.\\
We can now interpret the linearized equation~\eqref{eq:m-lin} as having a ferromagnetic and a spin-density wave contribution. Proximity to a SDW instability additionally lowers the critical interaction strength. Therefore, the metamagnetic transition could occur at a lower field then in the $z$-direction case, especially close to a SDW instability.\\
The magnetization due to a magnetic field applied in plane, as well as the amplitude of the staggered magnetization, are plotted in Fig.~\ref{fig:xy-meta}. Again, we find a first order transition for lower temperatures while the transition changes to a crossover upon increasing temperature. Note that the sign of $M^{y}$ depends on the sign of the spin-orbit coupling constant $\alpha$. There is no degeneracy in the state obtained which could lead to domain formation.\\
\subsection{Comparison}
\begin{figure}
	\centering
	\includegraphics{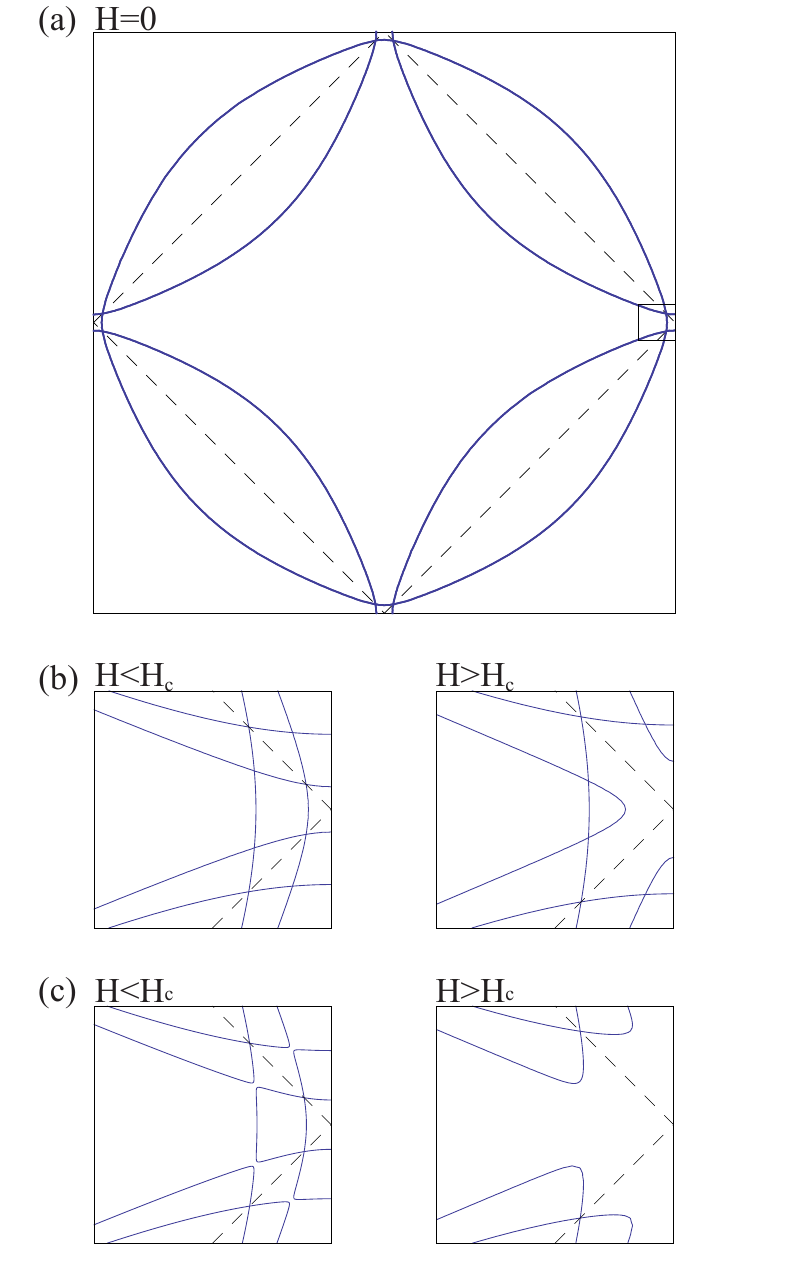}
	\caption{(a) Fermi surface for an electron density of $n=1.336$ without an applied field. Due to the rotation of the O-octahedra, the Brillouin zone is folded back, denoted by the dashed lines. (b) and (c): Fermi surface just below and above the critical fields in $z$- and $x$-direction, respectively. For clarity, only a small section of the BZ is shown indicated by the little square in (a). For the case of a field applied in plane (c), small gaps open close to the van Hove points.}
	\label{fig:fs}
\end{figure}
Comparing the two cases of fields applied in $z$- and $x$-direction in Figs.~\ref{fig:z-meta} and \ref{fig:xy-meta}, respectively, we first see that the critical field for the latter is shifted to lower fields, even though the zero-field susceptibilities differ by less then a percent, $\chi_0^z/\chi_0^x\approx0.998$. This is due to the proximity of the system to a SDW instability as was pointed out in the previous subsection.\\
This behavior is in qualitative agreement with the experimental phase diagram (see the schematic phase diagram in Fig.~\ref{fig:schem-phase}). However, the difference of the in-plane and out-of-plane critical field is smaller in our model calculation than in the experiment. Our model only includes the staggered spin-orbit coupling entering through the oxygen displacement. Naturally, other spin-orbit coupling contributions, particularly from the Ru-ions, would add to the anisotropy through an anisotropic $g$-tensor, likely with a larger polarizability in the basal plane than along the $z$-axis.~\cite{ng:2000b} This is, however, beyond the scope of this study as a detailed analysis would require to include other bands. \\
Second, a numerical study of the Gibb's free energy shows that the temperature $T^*$ up to which the first-order transition persists, is higher in the case of the in-plane field. For our choice of parameters we find $T_z^* \approx 9\cdot 10^{-4}t$ while $T_x^*\approx 11\cdot 10^{-4}t$. This anisotropy in the critical temperature is consistent  with the trend in the experimental situation. However, it does not reproduce the quantum critical endpoint. Note, that the difference between $ T^*_z$ and $ T^*_x$ could not be explained simply by an anisotropic $g$-tensor. In principle, it may be possible to tune the model in such a way as to press the critical temperature for out-of-plane fields, $ T^*_z $ to zero while still having a first-order transition at finite temperatures for in-plane fields. Also fluctuation effects are likely important in this context. These features are, however, not essential to our discussion.\\
An additional important difference between in-plane and out-of-plane fields can be seen in Fig.~\ref{fig:fs}, where the Fermi surfaces for both cases for fields below and above $H_c$ is shown: we see that the system undergoes a metamagnetic transition to prevent the majority-spin band from touching the van Hove points. In (c), we additionally see that the induced spin-density wave opens small gaps at the Fermi level close to the van Hove points. This has important consequences for the appearance of a nematic phase as we will see in the next section.

\begin{figure}
	\centering
	\includegraphics{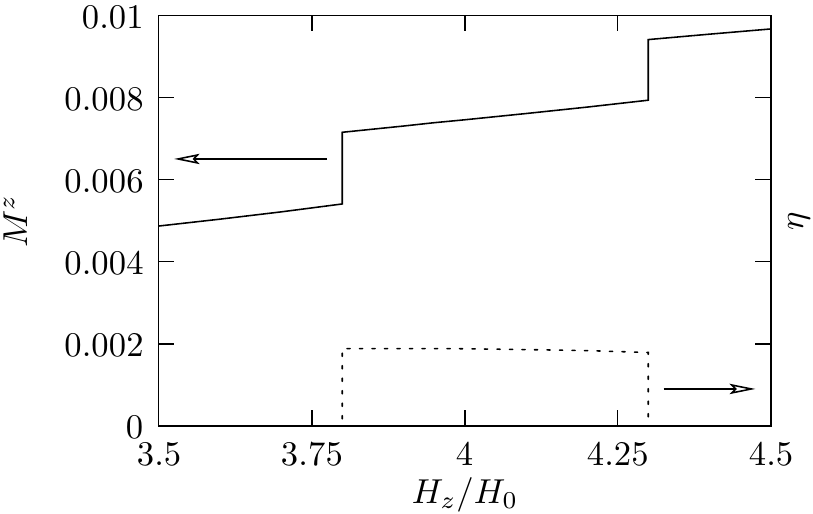}
	\caption{Mean-field results for the magnetization and the nematic order parameter $\eta$ for a field applied in $z$-direction for a temperature of $T=5\cdot10^{-4}t$ showing an intermediate nematic phase bounded by two first-order transitions. Here, $g = 0.33$ and $n=1.336$.}
	\label{fig:nem-z}
\end{figure}
\section{Nematic instability}
In this section, we explore the occurrence of a nematic phase in our model for the two cases of a magnetic field applied in $z$- and $x$-direction, respectively. For this purpose, we introduce an additional interaction term~\cite{yamase:2000c, metzner:2003}
\begin{equation}
	\HH^{n}=  \frac{1}{2 N}\sum_{\kk \kk'}\sum_{s s'}f_{\kk\kk'}n_{\kk s}n_{\kk' s'}
\end{equation}
with a coupling function $f_{\kk \kk'}$ only contributing for zero momentum transfer, i.e. for the forward scattering, which is the relevant interaction for a nematic phase to occur.~\cite{halboth:2000} We then separate the coupling function
\begin{equation}
	f_{\kk\kk'} = g d_{\kk} d_{\kk'}
\end{equation}
and choose a d$_{x^2-y^2}$ symmetric form for the form factors,  $d_{\kk} = \cos k_x - \cos k_y$. This term can then lead to a nematic phase, reducing the symmetry from $C_4$ to $C_2$.\\
Introducing again a mean-field decoupling which is spin-independent we write for this interaction
\begin{equation}
	\HH^{n}=\sum_{\kk s}\eta d_{\kk }n_{\kk s} - \frac{N}{2g}\eta^2
\end{equation}
with 
\begin{equation}
	\eta = \frac{g}{N}\sum_{\kk'}d_{\kk'}\langle n_{\kk'}\rangle.
\end{equation}
Since $d_{\kk} = - d_{\kk + \Q}$, but is isotropic in spin-space, we can deal with it by replacing
\begin{equation}
	\epsilon^{(1)}_{\kk} \rightarrow \tilde{\epsilon}^{(1)}_{\kk} = \epsilon^{(1)}_{\kk} + \eta d_{\kk}
\end{equation}
while all the above formulae still hold with the additional self-consistency equation
\begin{equation}
	\eta = - \frac{g}{N}\sum_{\alpha = 1,2}\sum_{\kk s} n_F(\xi_{\alpha\kk s})(-1)^{\alpha}\frac{d_{\kk}\tilde{\epsilon}^{(1)}_{\kk}}{ \sqrt{g_\textbf{k}^2 +(\tilde{\epsilon}^{(1)}_\textbf{k})^2}}
\end{equation}\label{eq:nem-z}
for the $z$-direction case and
\begin{equation}
	\eta = -\frac{g}{N}\!\sum_{\alpha, \beta}\!\!\sum_{\kk}n_F(\xi_{\alpha\beta,\kk}) \frac{(-1)^\alpha d_{\kk}(\tilde{\epsilon}^{(1)}_{\kk} \pm \tilde{h}_0^{x})}{\sqrt{\!(\tilde{m}^y\pm g_{\kk})^2\! +\! (\tilde{h}^x_0\pm\epsilon^{(1)}_{\kk})^2\!}}
\end{equation}
for the $x$-direction case, respectively.\\
\begin{figure}
	\centering
	\includegraphics{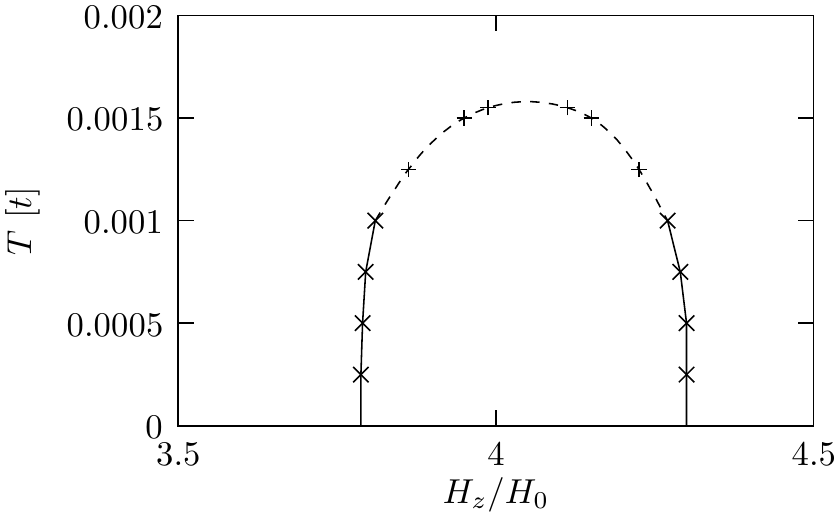}
	\caption{Phase diagram for a magnetic field applied in $z$-direction. While for low temperatures, the two consecutive transitions are of first-order (solid line), they become second-order before the nematic phase disappears completely (dashed line). Above this temperature, a metamagnetic crossover can still be seen.}
	\label{fig:phase-z}
\end{figure} 
For sufficiently strong $g$, we find a magnetization curve for fields applied in $z$-direction as is shown in Fig.~\ref{fig:nem-z}. The two jumps in the magnetization border an intermediate phase with a finite value of the nematic order parameter $\eta$. The instability in this case is again driven mainly by electrons whose momenta lie close to the van Hove points. To obtain an intermediate phase before a single metamagnetic jump removes all such electrons from the Fermi surface, a critical scattering strength $g_c$ is necessary. Above that, a nematic phase is entered at some magnetic field $H_{c1}$ and left again at $H_{c2}$.\\
The $T$-$H$-diagram shown in Fig.\ref{fig:phase-z} shows first-order transitions for low temperatures up to $T\approx0.001t$, second-order transitions for higher temperatures until at $T\approx0.0016t$ the nematic phase disappears completely to make way for a metamagnetic cross-over (not shown). This behavior has already been observed in similar calculations.~\cite{kee:2005, yamase:2007}\\
For the case of a field applied in $x$-direction, a very similar behavior is observed, however, with one important difference: Since the induced spin-density wave already removes some weight from the Fermi surface close to the van Hove points, a larger forward scattering strength is required for the occurrence of a nematic phase, i.e. $g_c^x > g_c^z$. Therefore, there is a range of $g$, where there exists already a nematic phase for fields in $z$-direction, but only one metamagnetic jump is observed for in-plane fields. This result is summarized in the phase diagram in Fig.~\ref{fig:nemz-nemxy}. As a function of the SOC strength $\alpha$ and the forward scattering strength $g$, we find three regions. In addition to the two obvious ones, where there is either no intermediate phase (region I) at all or one for fields applied in any direction (III), there is now a new region with a nematic phase only for fields applied in $z$-direction (II). Obviously, this region corresponds to the case of Sr$_3$Ru$_2$O$_7$.
\begin{figure}
	\centering
	\includegraphics{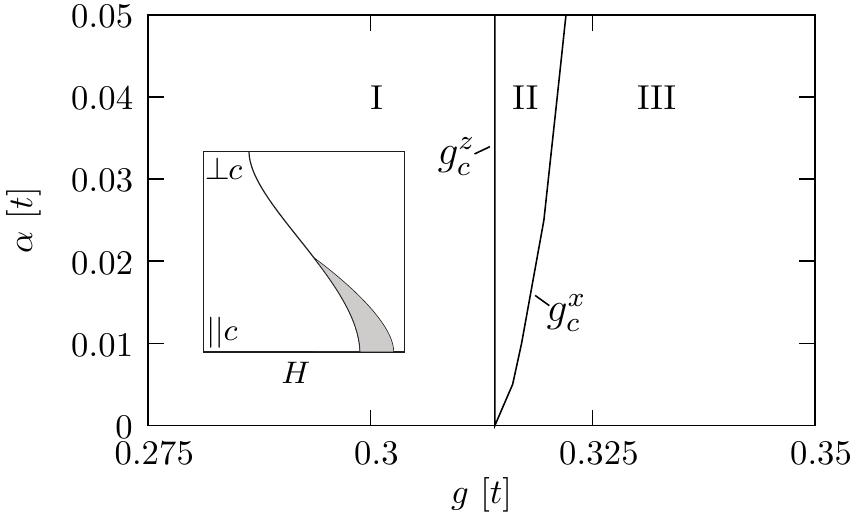}
	\caption{Critical forward scattering strength for the cases of an applied field in $z$- and $x$-direction, respectively. We can distinguish three different regions. I: The forward scattering strength is too week to enter a nematic phase, no matter in what direction the magnetic field is applied. Only a single metamagnetic transition occurs. II: While there is a nematic phase for fields applied in $z$-direction, no such phase occurs for fields in plane. This region corresponds to the case found in Sr$_{3}$Ru$_{2}$O$_{7}$. III: the forward scattering strength is strong enough such that a nematic phase will occur for fields in any direction. \emph{Inset}: schematic of the phase diagram showing that the nematic phase only occurs for fields close to the $z$-axis.}
	\label{fig:nemz-nemxy}
\end{figure}
\section{Discussion and Conclusions}
\begin{figure}
	\centering
	\includegraphics{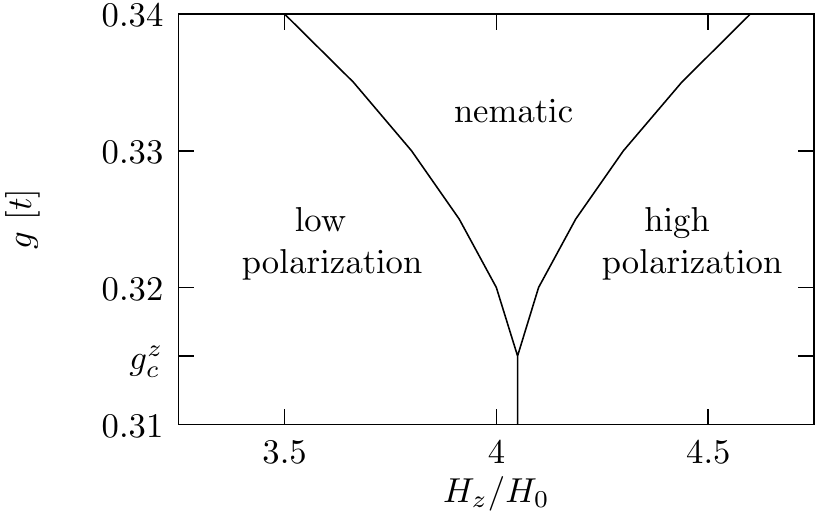}
	\caption{Phase diagram for a magnetic field applied in $z$-direction depending on the forward scattering strength $g$. Below a critical strength $g_c^z$, no nematic phase is entered, but the system undergoes a single metamagnetic transition connected to the proximity to the ferromagnetic instability as discussed above. Above $g_c^z$, the system enters a nematic phase whose region grows with increasing $g$. Here, $T=5\cdot10^{-4}t$.}
	\label{fig:g-H}
\end{figure}

The range of the forward scattering strength for which in our calculation a nematic phase is only observed for fields in $z$-direction is not very large. Note , however, that the nematic phase only appears in a very narrow region ($(H_{c2} - H_{c1}) / H_{c1} < 3\% $\cite{grigera:2004}) and thus, $g$ is only slightly bigger than $g_c^{z}$. This is illustrated in Fig.~\ref{fig:g-H} where the dependence of the critical fields on the forward scattering strength is shown. The width of the nematic phase grows rather rapidly with increasing $g$. Therefore, the actual size of the forward scattering strength might well lie in region II of Fig.~\ref{fig:nemz-nemxy}.\\
Obviously, not only the strength of the SOC, but also the nesting properties of the spin-polarized Fermi surfaces play an important role for the appearance of the anisotropy effect. Nesting properties are a factor of tuning our model to the vicinity of a SDW instability. To analyze the impact of enhanced SDW correlations, we consider two different electron densities and corresponding on-site interaction strengths $U$ with $U_c^{\rm FM} - U=$const (see crosses in the left part of Fig.6). This allows us to examine the cases of two different proximities to a SDW instability with comparable strengths of FM correlations. One important finding is that the stronger the SDW correlations the more pronounced the anisotropy effect and thus, the smaller the ratio $ H^x_{c} / H^z_{c} $  becomes.This is depicted in Fig.~14. In Fig.~6 we also show that the ferromagnetic and SDW instabilities can be competing for the chosen parameter range. We fix our model parameters in a way to avoid the occurrence of a staggered magnetic moment for any value of the magnetic field along the z-axis, while the staggered moment is field induced for in-plane fields.\\
We should also comment on the strength of the SOC $\alpha$ that we expect for this system. To get an estimate of the on-site SOC strength $\lambda$ for p-electrons on the oxygen, we take the O$^{2-}$ vacuum value, $\lambda\sim10$meV. A very crude estimate of the staggered SOC coupling from Eq.~\eqref{eq:hops} then yields
\begin{equation}
	\frac{\alpha}{t}\approx \frac{2 \lambda}{\Delta}\frac{t'_{pd}}{t_{pd}}\approx\frac{2 \lambda}{\Delta}.
\end{equation}
Taking estimates for $ \Delta \approx 1.5 eV $ we find that for $ \alpha $ a value on order of a percent of $t$ seems reasonable.~\cite{oguchi:95} Comparing Figs.~\ref{fig:nemz-nemxy} and \ref{fig:g-H}, this would allow for a nematic phase with a width of $(H_{c2}-H_{c1})/H_{c1}\approx 2\%$, in agreement with experiment. For a more reliable estimate of $\alpha$, DFT calculations should be performed.\\
Finally, some remarks to the nematic phase are in order. As was already mentioned, the nematic phase introduced here is the same as discussed by other authors.~\cite{kee:2005, yamase:2007b, doh:2007} As shown by these authors, the nematic phase  could account for several experimentally observed phenomena, like the anomalous resistivity, or the non-Fermi-liquid behavior of the susceptibility and the specific heat coefficient.\\
To conclude, we showed that the rotated oxygen octahedra lead to a staggered hopping that can be described with the help of a (staggered) spin-orbit coupling of Rashba-type. This introduces an anisotropy of the response to a magnetic field, namely an induced spin-density wave for the case of in-plane fields. This staggered magnetization could be observed in neutron scattering experiments. To our knowledge this kind of experiment has not been performed so far. The additional magnetization has, first, the effect that the critical field for a metamagnetic transition is shifted to lower values for in-plane fields. Also, the critical temperature $T^{*}$ up to which the transition is first order is higher for fields in the $xy$-plane. Last and most important, the spin-density wave opens gaps at the Fermi level that lead to an anisotropy for the appearance of a nematic phase.
Additionally considering spin-orbit coupling effects of the Ru orbital would account for the full anisotropy of $H_{c}$ ($g$-tensor anisotropy). Therefore, the present work allows for a picture that is qualitatively consistent with experimental observations including the anisotropies in $H_{c}$, $T^{*}$ and the appearance of a nematic phase.\\
Note added:  While preparing for submission we noticed the recent paper \cite{puetter:09} which studies the influence of spin-orbit coupling and the doubling of the unit cell on the nematic phase of Sr$_{3}$Ru$_{2}$O$_{7}$.
\section*{Acknowledgements}
We are grateful to  H. Adachi, B. Binz, F. Hassler, F. Loder, A. Mackenzie, Y. Maeno, T.M. Rice, A. R\"uegg, M. Ossadnik and  A. Thomann for helpful discussions. This work was financially supported by Swiss Nationalfonds and by the NCCR MaNEP.
\begin{figure}
	\centering
	\includegraphics{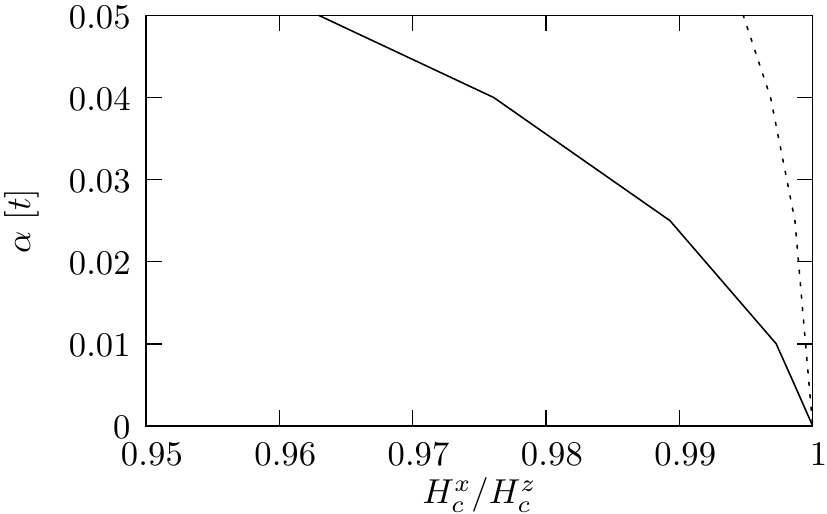}
	\caption{Anisotropy of the critical field in $x$-direction compared to the value for the $z$-direction for the densities $n=1.336$ (solid line) and $n= 1.339$ (dashed line), respectively. The value for the on-site interaction is chosen as indicated in Fig.~\ref{fig:Uc}. The dependence on the strength of the spin-orbit coupling is more pronounced the closer the system is to a SDW instability.}
	\label{fig:Hc}
\end{figure}

\end{document}